\documentclass[runningheads]{llncs}
\usepackage[T1]{fontenc}
\usepackage{graphicx}
\usepackage{amssymb}
\usepackage{comment}
\usepackage{rotating}
\usepackage{graphicx}
\usepackage{textcomp}
\usepackage{xcolor}
\usepackage[super]{nth}
\usepackage{multirow} 
\usepackage{colortbl}
\usepackage{nicematrix}
\usepackage{subcaption,booktabs}
\usepackage{float}
\usepackage[flushleft]{threeparttable}
\usepackage{algpseudocode}
\usepackage{makecell}
\usepackage{multicol}
\DeclareUnicodeCharacter{2212}{-}
\newcolumntype{L}[1]{>{\raggedright\arraybackslash}p{#1}}
\newcolumntype{C}[1]{>{\centering\arraybackslash}p{#1}}
\newcolumntype{R}[1]{>{\raggedleft\arraybackslash}p{#1}}

\begin{document}
\title{GXJoin: Generalized Cell Transformations \\for Explainable Joinability} %GXJoin: 
\author{Soroush Omidvartehrani%\inst{1}%\orcidID{0000-0002-3390-5194}
\and
Arash Dargahi Nobari%\inst{1}%\orcidID{1111-2222-3333-4444} 
\and
Davood Rafiei%\inst{1}%\orcidID{2222--3333-4444-5555}
}
\authorrunning{S. Omidvartehrani et al.} %F. Author et al.
\institute{University of Alberta, Canada \\
\email{\{s.omidvartehrani,dargahi,drafiei\}@ualberta.ca}}
\maketitle              % typeset the header of the contribution
\vspace{-7mm}
\begin{abstract} 
%Describing real-world entities can vary across different sources, posing a challenge when integrating or exchanging data. We study the problem of joinability under some transformations, where two columns are not equi-joinable but can become equi-joinable after some transformations. Discovering those transformations is a challenge because of the large space of possible transformations, which also grows with the input length and the number of rows. 
%Our focus is on the generality of transformations, aiming for broad applicability. We explore various generalization techniques, emphasizing those that yield transformations covering a larger number of rows and are often easier to explain. Recognizing that generalizing transformations can increase the search space, we introduce a method that significantly reduces the time needed for generating transformations. This is achieved by sampling rows containing more meaningful information for the model, as opposed to random sampling from the entire table.
%In our extensive evaluation on two real-world datasets, our approach demonstrates superior performance by generating fewer, and hence simpler and more explainable, transformations compared to a state-of-the-art approach. Importantly, this is achieved while effectively managing the search space and running time.
Describing real-world entities can vary across different sources, posing a challenge when integrating or exchanging data. We study the problem of joinability under syntactic transformations, where two columns are not equi-joinable but can become equi-joinable after some transformations. Discovering those transformations is a challenge because of the large space of possible candidates, which grows with the input length and the number of rows. Our focus is on the generality of transformations, aiming to make the relevant models applicable across various instances and domains. We explore a few generalization techniques, emphasizing those that yield transformations covering a larger number of rows and are often easier to explain. Through extensive evaluation on two real-world datasets and employing diverse metrics for measuring the coverage and simplicity of the transformations, our approach demonstrates superior performance over state-of-the-art approaches by generating fewer, simpler and hence more explainable transformations as well as improving the join performance. 

\keywords{Generalization \and Explainable Join \and Transformation}

\end{abstract}

\vspace{-2mm}
\section{Introduction}\label{sec:intro}
\vspace{-2mm}
Generalization is a fundamental concept across various branches of science. In the realm of machine learning, the notion of model generalizability serves as a measure of how effectively a model, learned from a limited set of training samples, can extend its performance to unseen data during testing. We study the problem of generalizability in transforming tables for joins, where the task is to transform data from one table formatting to another to make them equi-joinable, using only a limited sample of joinable row pairs. In this context, the transformations are used to explain and validate the join process and the generalizability of the transformations signifies the capacity to accurately represent the true mapping rather than capturing some form of \textit{accidental regularity} in the data. Also, a more general transformation  will garner more supporting evidence from the data and may be deemed more reliable, as opposed to a transformation that may capture random noise in the data. The problem of transforming tables has been explored in the literature~\cite{gulwani2011automating,gulwani2012spreadsheet,zhu2017auto,nobari2022efficiently}, with some variations in the supported set of transformations. However, the generality aspects of these transformations have not been investigated in these studies.

%In an era where data has become a valuable currency, we are no longer limited to internal data sources and we can benefit from external databases to access additional information that can impact our decisions. However, different stored data do not necessarily have the same format and querying and integrating them has always been a challenge. This task may range from merging two small tables to integrating a large amount of external data with an internal database when a company takes over its competitor. As we know, preprocessing various data from different sources in order to integrate them is a tedious and time-consuming task, which is not always possible using traditional techniques. 

%In this paper, we study the challenge of joining tables, where they are not equi-joinable. Specifically, our focus is on identifying general transformations that can map a column in one table to a column of another one with maximum coverage and simplicity in a reasonable time. These two columns must represent the same entities but with different formatting. One example is the columns of sources that store a name with different formattings, like \textit{Soroush Omidvartehrani}, \textit{S. Omidvartehrani}, and \textit{Omidvartehrani, Soroush}. Here, we are looking for transformations that can map one name to another. This transformation not only should be able to map a name from the source to the target column, but also we expect to cover other rows that are following a similar pattern. 

\begin{figure*}
  \centering
  \includegraphics[width=\linewidth]{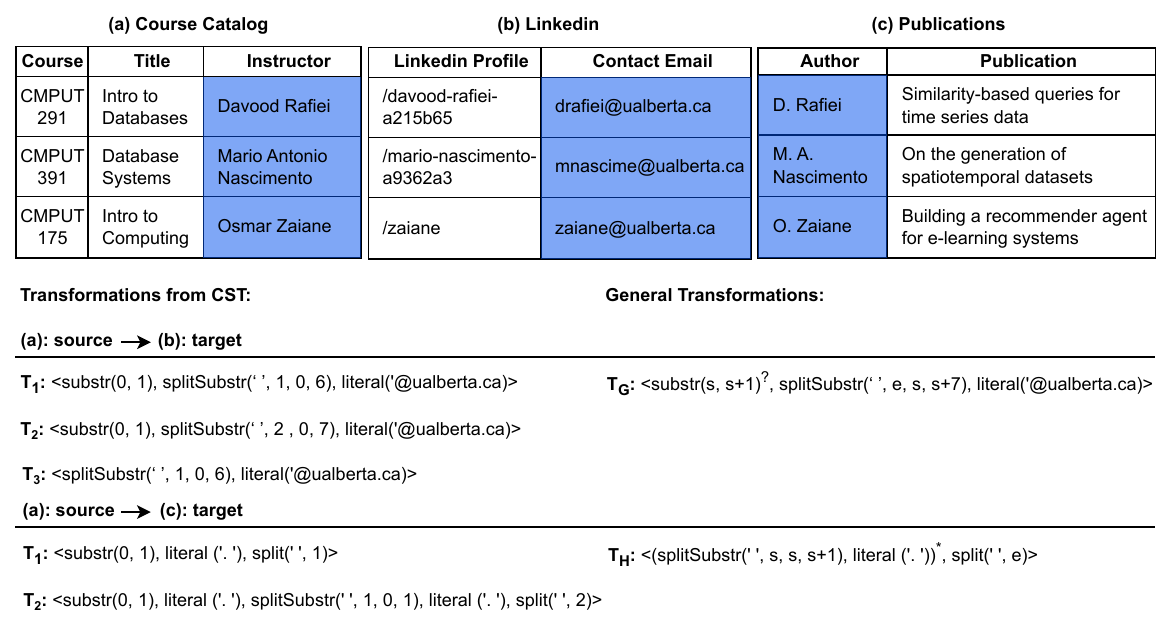}
  \caption{Example of joinable tables and transformations}
  \label{fig:main}
  \vspace{-6mm}
\end{figure*}

%\subsection{Motivating Example}
%In a university setting, a data analyst is tasked with the compilation of information about faculty members. The analyst has access to diverse data sources, encompassing university records, course catalogs, academic publications, and external sources. The overarching objective for the analyst is to seamlessly integrate data from these disparate sources, constructing a comprehensive profile of professors, their courses, and research contributions. This aggregated data holds the potential to offer a holistic perspective on faculty members, facilitating program promotions, efficient course planning, strategic resource allocation, faculty evaluations (e.g., tenure cases), and enabling further in-depth analysis.

\paragraph{\textbf{Motivating Example:}} A data analyst is tasked to integrate information about faculty members from various sources to create comprehensive profiles.
Consider the tables presented in Figure~\ref{fig:main} from three different sources. Despite describing the same entities, these tables are not equi-joinable due to formatting mismatches. The manual construction of a mapping between these tables can be both time-consuming and may necessitate domain knowledge that the analyst might not possess.
Suppose the analyst aims to map the instructor name in Table (a) to the contact email in (b) and the author name in (c). Leveraging a recent work on transforming tables~\cite{nobari2022efficiently}, referred to as CST and regarded as state-of-the-art for its improvement upon previous works, the analyst obtains three transformations, denoted as $T_1$, $T_2$, and $T_3$ to transform the instructor name in Table (a) to email address in (b), as shown in the Figure. $T_1$ caters to the first row, extracting the first character of the first name, concatenating it with the last name, and adding a literal to generate the email address in the target table. $T_2$ follows a similar approach, but accommodates the second row, which includes a middle name. $T_3$ constructs the email based on the last name only and is applicable to the third row. However, with each transformation applicable to only a single row, the generality of them can be questioned. On the other hand, the transformation $T_G$, which utilizes optional fields and relative indexes, works universally for all three rows, offering a higher level of generality. One could argue that $T_G$ better captures the underlying pattern and is more readable. The same rationale applies to the mapping of (a) to (c), where CST identifies two transformations to map the tables, whereas a more general transformation, denoted as $T_H$, exists, covering all table rows\footnote{The transformation functions and generalization methods are further explained in Section~\ref{sec:approach}.}.

%\subsection{Problem Studied}
%In general, we prefer more general transformations that cover a large number of input rows, as such transformations are more likely to cover unseen data. Moreover, these transformations may better show the underlying patterns in the data, making it easier for a data scientist to verify.
%However, the pursuit of more general transformation raises a few intriguing questions that have not been well-explored:
%(1)~What principles of generality can be applied to transformations, and are these generalizations effective when applied to tabular data?
%(2)~What are the overheads or costs associated with generalizing transformations?
%(3)~Can these overheads be circumvented using more efficient approaches?
%A significant challenge arises from the vast space of possible transformations and their parameters, with the search for more general transformations further amplifying the size of the search space. 
%We study these problems in the context of transforming tables for explainable joinability.

\paragraph{\textbf{Problem Studied:}} We prefer more general transformations that cover a larger number of input rows, as they are more likely to cover unseen data, may better show the underlying patterns in the data, and make it easier for a data scientist to validate the join process. However, there has been limited exploration into which principles of generality can be applied to transformations and whether these generalizations are effective when applied to tabular data. We study this problem in the context of transforming tables for explainable joinability, with the joined columns typically describing the same entities.

\vspace{-3mm}
\paragraph{\textbf{Our Approach:}} Our work aligns with the line of work on program synthesis, aiming to discover cell transformations that enable an equality join. It also  falls within the realm of \textit{explainable join}, where transformations explain the reasons behind a join~\cite{zhu2017auto,nobari2022efficiently}, albeit the discovered transformations may also aid in predicting missing values~\cite{gulwani2011automating,singh2016blinkfill}, conducting schema mapping~\cite{cate2017approximation,bonifati2019interactive}, and performing data cleaning and repairing~\cite{he2016interactive}. Our work explores several generalization concepts in transforming tables, incorporating those observed in our motivating example, and explores their impact on reducing the number of transformations while maintaining or improving their coverage. Our evaluation on two real-world datasets obtained from different sources reveals that generalization techniques significantly enhance transformation coverage, reduce the number of transformations required to map all rows, better address unseen data, and enhance the end-to-end result of the joinability task.
Our contributions can be summarized as follows: (1)~We explore the impact of generalization on transforming tables for explainable joinability, aiming to improve the adaptability and coverage of transformations. (2)~We expand the domain of transformations by incorporating generalization concepts, including length invariance, repeating patterns, and bi-directionality. (3)~Our comprehensive experiments unveil the superior performance of our method, compared to a state-of-the-art approach, across various metrics, including the coverage of transformations, the number of required transformations for mapping, and the ability to cover previously unseen data.\footnote{Our code and the datasets are available at \url{github.com/soroushomidvar/GXJoin}}

\vspace{-2mm}
\section{Related Works}\label{sec:rel}
\vspace{-2mm}
The pertinent works related to ours can be categorized into (1) finding related tables and joinable row pairs, (2) discovering example-driven table transformations to enable an equi-join, and (3) model generalization.

\noindent
\textbf{(1) Finding Related Tables and Joinable Row Pairs:} 
When data to be integrated are located in a data lake, a common initial step involves identifying related tables and potential joinable columns. Finding related tables is studied within the context of joinable tables~\cite{josie,zhu2016lsh,WebTables}, unionable tables~\cite{Table_union2018Nargesian,ling2013synthesizing}, and tables that are semantically similar~\cite{table2vec,zhang2021semantic}. 
%Some of these approaches can be leveraged within our framework to retrieve candidate tables. For instance, Zhu et al.~\cite{zhu2016lsh} address the problem of efficiently identifying columns from the same domain in a large dataset. Their approach involves finding columns that share many common values with a seed column, utilizing LSH indexes~\cite{zhu2016lsh} to scale up the lookup process and a novel data partitioning method optimized for LSH indexes. 
With relevant tables retrieved, many approaches, including ours, rely on a set of provided examples--whether human-provided or automatically generated--to perform a tabular join through some transformations. 
%In the latter case, the accurate identification of matching rows to serve as examples becomes crucial. 
The problem of finding joinable row pairs where the matching pairs are not equi-joinable is well-studied~\cite{Chaudhuri2006fuzzy,fastjoin,auto-em,wang2014extending}, with methods ranging from basic textual similarity functions~\cite{Chaudhuri2006fuzzy} to bipartite graphs of tokens and their similarity~\cite{fastjoin,wang2014extending}, and deep learning-based entity matching techniques~\cite{auto-em,Li2020:Deep}. While the predominant focus of this line of work revolves around matching relevant rows, our approach is centered on the generalization of explainable transformations that render two rows equi-joinable. Hence these approaches can serve as preprocessors in our approach, aiding in the generation of examples.
% ~\cite{Chaudhuri2003Robust,Chaudhuri2006fuzzy,fastjoin,MassJoin,yu2016string,auto-em,wang2014extending,MFJoin}

\noindent
\textbf{(2) Discovering Example-Driven Table Transformations:}
The transformation of cells within a table is recognized as a crucial operation in tabular data integration~\cite{openRefine,heer2015predictive}, and the discovery of such transformations, guided by a set of examples, has been the subject of several studies~\cite{nobari2022efficiently,zhu2017auto,singh2016blinkfill,gulwani2011automating,gulwani2012spreadsheet}.  FlashFill~\cite{gulwani2011automating,gulwani2012spreadsheet} and BlinkFill~\cite{singh2016blinkfill}, the two pioneering contributions in this area, create an input graph based on user-provided examples and traverse the graph to generate transformations. Nevertheless, they heavily rely on accurate input examples and struggle when the examples are noisy. 
The huge search space and resource-intensive computation in FlashFill have inspired the subsequent works to prioritize a smaller search space over generalizability. 
For instance, Auto-join~\cite{zhu2017auto} employ predefined string-based units to generate transformations and a recursive backtracking algorithm to find an optimal one. Limiting the search space to a predefined set of transformations allows Auto-join to handle minor input noise. %, but it resorts to a backtracking algorithm that becomes computationally expensive when iterating through the entire transformation space. 
In a more recent study, Common String-based Transformer (CST)~\cite{nobari2022efficiently} constrains the search space by utilizing common text sequences between source and target examples as textual evidence to construct the skeleton of transformations. 
%Similar to Auto-join, CST employs string-based transformation units but generates transformations for each row independently, enhancing noise handling in the input. The resulting transformations are ranked based on their coverage to compile a final transformation set. CST demonstrates improved runtime performance and noise handling compared to Auto-join. 	
 The primary focus of all the aforementioned approaches has been on identifying transformations that cover a given example set, with less focus on the generalizability of the transformations to newly added data and unseen examples. %This limitation has been a significant drawback in previous studies and serves as the primary focus area of this work. 

%Despite the improved runtime and search space reduction CST, it may miss certain transformations, especially those not covered by the small set of predefined transformation units. To overcome this issue, Nobari et al.~\cite{dtt}, in a work referred to as Deep Tabular Transformer (DTT), exploited Large Language Models (LLMs) and developed a deep learning framework to transform source into target. DTT is claimed to noticeably outperform the aforementioned approaches in terms of accuracy and supported transformation classes. 

%In this paper, we focus on the generalizability of the generated transformations and form up the units in a way that better complies with gradual data changes and new entries.

\noindent
\textbf{(3) Model Generalization: }
Within the broader context of our work, there exists a substantial body of literature on generalizing models for diverse inputs. Improving the generalization capacity of neural architectures, often confined to particular domains and tasks, is a well-studied topic~\cite{russin2019compositional,herzig2020span,chen2020compositional,liu2020compositional}. In a closely aligned work, Shi et al.~\cite{shi2022compositional} study the compositional generalizability of transformer models, where they identify high-level categories of generalization, such as length generalization and mixing and matching of artifacts to improve the capacity of transformer models trained on simple subtasks to solve more complex tasks.
% The authors analyze the effectiveness of program synthesizers to generalize and the capacity of transformer models trained on simple subtasks to solve more complex tasks. 
% Additionally, data augmentation has been employed as a generalization approach~\cite{wang2021learning,akyurek2020learning,oren2021finding}. By augmenting the data, models are trained on a wider range of data variations, learning to capture the underlying patterns and reducing the likelihood of memorizing specific examples. 
%~\cite{qiu2021improving,wang2021learning,akyurek2020learning,oren2021finding}
%Some studies demonstrate that datasets augmented with recombined and reshaped examples yield better domain generalization than datasets with carefully constructed examples~\cite{andreas2019good,raileanu2020automatic,volpi2018generalizing,wu2020generalization}. 
Moreover, There is also a growing interest in meta-learning, enabling models to adapt to new tasks or domains by leveraging knowledge acquired from previous tasks. Some recent works~\cite{conklin2021meta,wang2020meta,lake2019compositional} identify abstract features that prove useful across different problem domains, hence enhancing generalization capabilities.
To the best of our knowledge, the generalization of cell transformations for explainable joinability and its impact on performance have not been previously explored.

\vspace{-2mm}
\section{Preliminaries}\label{sec:preliminaries}
\vspace{-2mm}

% In this section, we provide the necessary foundation for understanding our subsequent discussions, including the end-to-end process of an explainable join and the definition of cell transformations within the context of unit-based mapping.

In this section, we discuss the necessary foundation for our framework and the methods that our approach builds upon. 

%As depicted in Figure~\ref{fig:end-to-end}, our work is part of a larger data discovery and integration process, which typically begins with identifying relevant tables and joinable columns, proceeds to finding joinable row pairs, and culminates in generating explainable transformations for matching rows. 
%Generalization techniques may be employed either during or after the transformation generation phase and prior to utilizing these transformations for joining the tables.
%Our approach involves applying our generalization strategies, which may be employed either during or after the transformation generation phase, to a base set of transformations and evaluating the impact on the coverage of the input and the overall quality of the transformations. 
%For our base transformations, we utilize the set introduced in recent relevant literature (as discussed next) and extend them using the generalization strategies. 

\vspace{-2mm}
\subsection{An End-to-End Explainable Join Algorithm}
\vspace{-2mm}

% As depicted in Figure~\ref{fig:end-to-end}, our work is part of a larger data discovery and integration process, which typically begins with identifying relevant tables and joinable columns, and proceeds to finding joinable row pairs. The row pairs serve as input in a typical unit-based transformation generation model to construct transformation skeletons and generate all possible transformations. The transformations are then sorted based on their coverage and top k transformations that can map all rows will be selected. In this process, our approach involves applying the generalization strategies, which may be employed either during or after the transformation generation phase, to a base set of transformations and evaluating the impact on the coverage of the input and the overall quality of the transformations. 

\begin{figure*}[t]
  \centering
  \includegraphics[width=\linewidth]{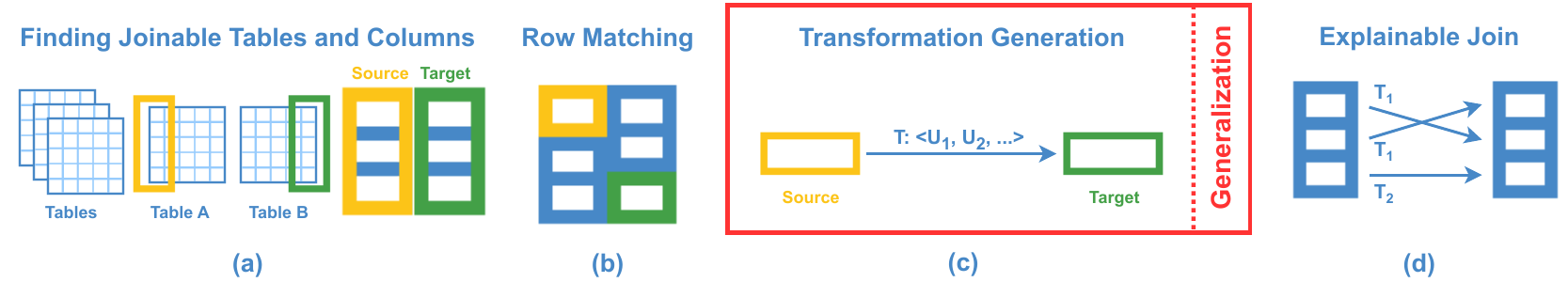}
  \caption{The end-to-end process of an explainable join}
  \label{fig:end-to-end}
  \vspace{-5mm}
\end{figure*}

Figure~\ref{fig:end-to-end} depicts an overview of a typical unequal join framework. The process begins by identifying relevant tables and joinable columns, followed by the formation of source-target pairs for potentially joinable rows. In some cases, data scientists may provide example row pairs. String-based example-driven transformers such as CST~\cite{nobari2022efficiently} utilize the provided example set to construct transformation skeletons and generate a large set of candidate transformations along with their coverage. With candidate transformations and their coverage known, a transformation may be selected based on its coverage, or a minimal set of transformations covering all given example row pairs may be sought. Our work is on the generalization of the framework, where some strategies are applied either during or after the transformation generation phase, evolving them into more general ones, which can lead to better coverage and a reduced set of covering transformations.

\vspace{-2mm}
\subsection{Cell Transformations}\label{sec:main-background}
\vspace{-2mm}
A \textit{cell transformation} (in short, a transformation) is a function that takes an input cell value represented in the source formatting and generates the representation of that value in a desired target formatting. 
% A large number of such transformations may be applied to table cells\footnote{ \url{https://openrefine.org/}}. 
We follow a recent line of work relying on cell transformations to enable unequal joins~\cite{zhu2017auto,nobari2022efficiently} where cell transformations are constructed as a sequence of basic operations referred to as \textit{units}. Each unit is applied on a cell value in the source and generates a fraction of a value in the target. The outputs of multiple units may then be concatenated to form the full cell value in the target. The units are string-based functions that either copy part of the input or a constant literal to the output.
A wide range of functions have been employed as units in the literature~\cite{zhu2017auto,nobari2022efficiently,gulwani2012spreadsheet,singh2016blinkfill} and our work adopts four commonly occurring units from previous studies~\cite{zhu2017auto,nobari2022efficiently}:
\begin{itemize}
\item \textbf{literal($str$)} returns $str$ as the output,
\item \textbf{substr ($i$, $j$)} provides the substring starting at index $i$ and ending at index $j$ or the end of the input if the input length is shorter,
\item \textbf{split($c$, $t$)} tokenizes the input based on character $c$ and returns the $t^{th}$ token, 
\item \textbf{splitSubstr($c$, $t$, $i$, $j$)} combines split and substring functions, with the input tokenized using character $c$ and subsequently substr($i$, $j$) is applied on the $t^{th}$ token. 
\end{itemize}
One may question the significance of the splitSubstr unit when split and substr units already exist individually. The reason for these compound units is that transformations are formed as a concatenation of units, and stacking of the units is only possible by introducing new units (e.g., splitSubstr).

For a given set of source and target pairs $(s_i, t_i)$, the coverage of a transformation is defined as the fraction of pairs in which applying the transformation on $s_i$ generates $t_i$. When a transformation can produce more than one output, the coverage is defined as the fraction of pairs in which applying the transformation on $s_i$ leads to a set that includes $t_i$. %The coverage of a transformation plays an important role in the joining process.
When transformations are employed to enable a join between two tables that are not otherwise equi-joinable, each transformation is applied to a row in the source, producing one or more candidates that are then looked up in the target.

\vspace{-2mm}
\section{Methodology}\label{sec:approach}
\vspace{-2mm}

% Our approach involves applying our generalization strategies (as discussed next) to a base set of transformations and evaluating the impact on the coverage and the overall quality of the transformations.

In this section, we introduce our generalization strategies and discuss how they are applied to a base set of transformations. We also analyze their impact on the coverage and the overall quality of the transformations.

%Our approach involves applying our generalization strategies to a base set of transformations and evaluating the impact on the coverage of the input and the overall quality of the transformations. For our base transformations, we utilize the set introduced in recent relevant literature (as discussed next) and extend them using the generalization strategies. 
%($\S$~\ref{sec:generalization-strategies}). 
%This process has implications for the search space and running time. We address those issues with our cluster sampling ($\S$~\ref{sec:sample}).

\vspace{-2mm}
\subsection{Length Generalization Using Relative Indices}
\vspace{-2mm}
The invariance of a table transformation to input length is a desirable property, as it is expected to mitigate overfitting, especially if the input length varies between the extracted examples given as context and those used for testing the transformations. 
Many parameters in a transformation make references to input using some indices, and those indices can be absolute or relative. Relative indices need an anchor point, which may be set to the beginning, the end or any other input position.
All transformations introduced in Auto-join and CST use absolute indices. Those indices can be replaced with relative indices without changing their semantics if we set the anchor point the beginning of the input. However, one can change the anchor point of a relative index to a different location, allowing the expression of transformations that will not be possible otherwise. 
%With more anchor points, the chance that two arbitrary rows will have the same indices under at least one anchor point is higher.

As possible anchor points, let $s$ and $e$ mark the beginning and the end of input respectively. Relative indices in a transformation can be stated with respect to either $s$ or $e$, allowing two-way parsing of the input.    
For example, consider the first and second rows from the Course Catalog and Linkedin tables in our running example from Section~\ref{sec:intro}, as depicted in Figure~\ref{fig:main}. The transformation \textit{<substr(0, 1), splitSubstr(‘ ’, 1, 0, 6), literal(`@ualberta.ca')>}, generated under CST,  maps the first row but not the second because of the presence of a middle name. However, if we use relative indices, the transformation \textit{<substr(s, s+1), splitSubstr(‘ ’, e, s, s+7), literal(`@ualberta.ca')>}, generated from the second row, will cover both rows.

\vspace{-2mm}
\subsection{Recurrence Generalization}% : Unit Repetition} %\subsection{Unit Repetition}
\vspace{-2mm}
Systematically handling repeating patterns is a generalization strategy that can reduce the complexity of integrating diverse data sources, improve the simplicity of mapping rules and potentially decrease the likelihood of errors in the integration process.
In the context of table transformations, a unit that is applicable to an element in an input sequence may be applicable to all elements in the sequence.
%Some patterns repeat in the input, and the mechanism for mapping them may also be the same. 
For example, consider a sequence of space-separated terms in a source column and a transformation that takes the first letter of each term and concatenates them together to produce the target. If the input length, in terms of the number of terms, is fixed at $k$ for all rows, one can construct a sequence of $k$ splitSubstr functions, each taking a term in the input sequence and outputting the first character of the term. However, this approach will not work if the input length varies from one row to the next. 
%The alternative of writing a transformation for each input length is too cumbersome, resulting in less clean and readable mapping.
%By introducing repetition into our transformation rules, we aim to account for recurring events and ensure that similar event sequences are treated consistently. 
We introduce repeating patterns, in the form of unit repetition and unit removal, to address this problem.

\vspace{-4mm}
\subsubsection{Unit Repetition}
Introducing repetition to transformation rules is not as straightforward as one may wish. This is because our transformation units use indices to indicate parts of input they operate on, and the behavior of those indices must be defined under repetition.
For example, consider \textit{split(` ', s)} which breaks the input based on the space character and takes the first element. Repeating this operation without changing the index will produce the same output and is less desirable.
Since we want to break the input into smaller segments and operate on each segment through repetition, we will focus on generalizing the \textit{split} and \textit{splitSubstr} functions\footnote{\textit{splitSubstr} becomes \textit{substr} when the splitting character does not exist in the input, hence \textit{substr} is not considered.}.
 
\begin{definition}
We define \textit{split} and \textit{splitSubstr} under repetition as   
\[
split(c,i)^r = \{split(c,i+j) | j \in [0,r-1]\},
\]
\[
splitSubstr(c,i,p,q)^r = \{splitSubstr(c,i+j,p,q) | j \in [0,r-1]\},
\]
where $r$ is the repetition factor, $c$ is separator character(s) breaking the input into segments, and $i$, $p$, and $q$ are indices, which can be both relative and absolute indices, and $i+j$ is constrained by the input length.
\end{definition}
The concept of repetition can be applied to either a single unit or to a sequence of consecutive units within a transformation. 
As an example, consider the first and second rows from the Course Catalog and Publications tables in Figure~\ref{fig:main}.
The transformation \textit{<splitSubstr(` ', s, s, s+1), literal (‘. ’), split(‘ ’, e)>} maps the first row of the source table to the first row in the target table, but it cannot map the second row because of the presence of a middle name. However, the first two units can be applied multiple times under repetition. In particular, with at most two repetition of those two units, both rows are covered.

\subsubsection{Unit Removal}
As a special case of repetition, both being or not being a unit in the rule can generate valid outputs. 
% a unit may or may not be applied. In other words, both being or not being a unit in the rule can generate valid outputs. 

\begin{definition}
We define \textit{split}, \textit{substr}, and \textit{splitSubstr} under removal as   
\[
split(c,i)^? = split(c,i) | \varnothing,
\]
\[
substr(p,q)^? = substr(p,q) | \varnothing,
\]
\[
splitSubstr(c,i,p,q)^? = splitSubstr(c,i,p,q) | \varnothing,
\]
where $?$ represents the removal concept, and the separator character and indices remain as previously defined.
\end{definition}

%Figure~\ref{fig:rem} illustrates this concept, where transforming the second row becomes possible by removing a subpart of a previously generated transformation. 

The transformation $T_G$
%\textit{<$substr(s, s+1)^?$, splitSubstr(` ', e, s, s+7), literal(`@ualberta.ca')>} 
applied to the second and third rows from the Course Catalog and Linkedin tables in Figure~\ref{fig:main} illustrates this concept, where transforming the third row becomes possible by removing a subpart of a previously generated transformation. 

\begin{comment}
\begin{figure*}[t]
  \centering
  \includegraphics[width=0.95\linewidth]{fig/rem.pdf}
  \caption{Example of unit removal concept}
  \label{fig:rem}
\end{figure*}
\end{comment}

\begin{comment}

\subsection{Changing Unit Order}
Altering the orders of units in a rule can result in generating different outputs. This manipulation can help to build another meaningful rule based on an existing one. An example of changing unit orders can be seen in Figure~\ref{fig:ord}, where the generated transformation for the first row returns the unit number at the end but by changing the order of literal and split functions, it can also cover the second row. 

\begin{figure*}
  \centering
  \includegraphics[width=0.95\linewidth]{fig/ord.pdf}
  \caption{Example of changing unit order concept}
  \label{fig:ord}
\end{figure*}

\end{comment}

\vspace{-2mm}
\subsection{Source-Target Direction Generalization}\label{sec:bidirec-gen}
\vspace{-2mm}
When dealing with numerous tables from different sources, determining which table should serve as the source and which as the target in the mapping process can be challenging. Ideally, we prefer a more content-rich table to act as the source, providing ample evidence to construct the target. However, identifying such content-rich tables may not be straightforward.
As an example, consider Course Catalog and Linkedin tables in Figure~\ref{fig:main}. One heuristic for detecting the source involves selecting the more informative column based on length~\cite{nobari2022efficiently}. This implies choosing the column with longer text as the source and treating the column with shorter text as the target. While this strategy may work for some table pairs, it is evident that it would fail in selecting the source here. In this case, the email column is longer due to the literal at the end, causing it to be chosen as the source, whereas the name column is more descriptive.

One potential generalization is to avoid fixing the source and target and allow mapping in both directions. One can dynamically choose the direction that yields more general transformations. In our case, we take a small subset of each table as a sample and generate transformations based on this sample for both directions. The direction that can map rows with fewer transformations is deemed as the source, as those transformations are expected to exhibit better generalization. 
%The impact of sampling and its effectiveness in discovering transformations will be discussed in detail in Section \ref{sec:performance}.
%To label the source and target correctly, we propose to sample a small subset of each table pair and generate transformations from both directions. The direction that can map rows with fewer transformations will be considered as the source (and vice versa) because this direction is able to generate more meaningful transformations. The effect of sampling and its ability to discover transformation will be discussed in Section \ref{sec:performance} in detail. 

\vspace{-2mm}
\subsection{Simplicity Generalization on Unseen Data}\label{sec:simplicity}
\vspace{-2mm}

In many real-world scenarios, table rows undergo updates, and new rows are regularly added over time. In such dynamic environments, there is a need to continuously incorporate new data in the join process. However, repeatedly generating transformations from scratch each time new data arrives can be a time-consuming and resource-intensive task. A general transformation is expected to not only cover the existing rows, but also anticipate unseen rows that may arrive in the future. One approach to gauge the generality of a transformation is to estimate its coverage based on data observed thus far. A transformation that covers more rows is deemed more general. However, without comprehensive knowledge about the distribution of future data, it is generally challenging to say with certainty whether one transformation will offer greater coverage than another.
A significant issue impacting the generality of transformations is the existence of accidental patterns in data. While such patterns may be present in a small sample, they are less likely to generalize to a larger sample or unseen data. 
Consequently, transformations constructed using a small sample may exhibit some of these accidental patterns. 
To mitigate such occurrences, we introduce simplicity as an additional measure of generality. In this context, simpler transformations are favored over more complex ones when all other factors, such as coverage on seen data, remain the same.
In our case, a simplicity score may be defined in terms of either the number of units or the number of parameters in a transformation, where a transformation with fewer units or parameters is considered simpler. In our experiments, transformations are primarily chosen based on their coverage. In cases where there is a tie in coverage, a simpler transformation is favored over a more complex one. For example, consider two generated transformations with identical coverage on input data, where all their units are the same except for one. Suppose one transformation has a \textit{substr(i, j)} unit, while the other includes \textit{splitSubstr(c, 0, i, j)}. When  character \textit{c} is absent from the input data, both transformations have the same coverage. However, when considering transformation simplicity, we favor the former because if unseen data has a matching character \textit{c} by chance, only \textit{substr(i, j)} may  perform the correct mapping.

\vspace{-2mm}
\section{Experiments}\label{sec:experiments}
\vspace{-2mm}
%In this section, we present the experimental results assessing the effectiveness and performance of our proposed method. 

Our evaluation is conducted on two real-world datasets gathered from diverse sources with varying levels of noise and inconsistency, 
%, which are sourced from diverse origins and characterized by different data distributions and patterns, 
providing opportunities to evaluate various aspects of our generalization techniques. 
%Before reporting the results and analysis, we present the datasets and the setup used in our experiments.
\textit{\textbf{(1)~Web Dataset}}~\cite{zhu2017auto} is generated by sampling table-related queries from the Bing search engine and retrieving sets of tables from Google Tables corresponding to those queries. The tables in this dataset are manually selected as those representing the same entities with different formatting that are joinable via textual transformations. This dataset consists of 31 pairs of web tables, covering 17 diverse topics. On average, each table contains approximately 92 rows, and the average length of a join entry is 31 characters. It is a challenging benchmark due to inconsistencies in the data and the presence of different formatting and patterns across the rows of each key column. \textit{\textbf{(2)~Spreadsheet Dataset}}~\cite{alur2016sygus} is taken from the public transformation benchmarks in FlashFill~\cite{gulwani2011automating,gulwani2012spreadsheet} and BlinkFill~\cite{singh2016blinkfill}. This benchmark is formed by the data cleaning challenges reported by the spreadsheet users on Microsoft Excel forums. The dataset contains 108 table pairs, each with 34 rows and 19 characters per join entry on average.

In our experiments, we set the maximum number of units and repetition degree to $3$ and $2$ respectively. We assume a manually annotated set of joinable row pairs is provided, as finding the matching pairs is beyond the scope of this study. 
% There are some approaches in the literature for finding such pairs when they are not given~\cite{nobari2022efficiently}. 
% We exploit the manually annotated joinable row pairs that are available in both benchmarks as our source-target examples. 
Moreover, coverage values are averaged across all tables in the benchmark, and in cases where randomness affects the results, experiments are repeated at least 5 times, and the average is reported.
%, unless explicitly stated otherwise. 
Our baseline for comparison is CST~\cite{nobari2022efficiently}, which, to the best of our knowledge, represents the state-of-the-art in finding explainable textual transformation. 

\vspace{-2mm}
\subsection{Transformation Coverage}
\vspace{-2mm}
%In this section, we study the impact of our proposed generalization techniques on the coverage of generated transformations, as compared with the baseline approach.

%\subsubsection{Best Transformation Coverage}
In this experiment, our aim is to generate transformations that cover all rows in the input. Since the framework is designed to generate a covering set, the coverage in our experiments is 100\%, for both our approach and the baseline. To evaluate the impact of our generalization, we introduce two alternative metrics to measure the coverage of transformations: 
(1)~\textit{best transformation coverage}, which indicates the coverage achieved by the transformation that produces the highest coverage on the input rows; and (2)~\textit{the number of generated transformations}, because, when a smaller number of transformations is required to cover all input rows, it serves as an indicator that the transformations are more general.
Table~\ref{tab:bestcov-a} summarizes the best transformation coverage when applying each generalization technique, as well as that of the baseline approach. 
%The reported values are the average over all tables in each dataset. 
Table~\ref{tab:bestcov-b} demonstrates the same evaluation, excluding tables that are covered by a single transformation. These excluded tables are considered easy, as a single transformation with non-general units may also achieve full coverage, leaving no room for generalization. Finally, Table~\ref{tab:reqnum} shows the total number of generated transformations required to achieve a coverage of 1.00 on all tables in the benchmark. The numbers inside parentheses indicate the improvement compared to the baseline. We analyze the effect of each generalization strategy individually:

\begin{table}[t]
\tiny
\caption{Best transformation coverage}
\centering
\begin{subtable}{0.495\textwidth}
\centering\centering{\bf
\begin{tabular}{|C{9mm}|C{9mm}C{8mm}C{9mm}C{9mm}C{10.5mm}|}
\rowcolor[HTML]{89A8F5} 
\hline
Dataset     & Baseline & Bidir.         & Rel.           & +Rem.           & + Rep.       \\ \hline
Web         & 0.577    & 0.606 (2.90\%) & 0.676 (9.90\%) & 0.679 (10.21\%) & \makecell[tc]{0.679 \\ (10.21\%)} \\ \hline
Spr. & 0.667    & 0.667 (0.00\%) & 0.708 (4.09\%) & 0.725 (5.79\%)  & \makecell[tc]{0.725 \\ (5.79\%)}  \\ \hline
\end{tabular}}
\caption{On all tables} \label{tab:bestcov-a}
\end{subtable}
\begin{subtable}{0.495\textwidth}
\centering{\bf
\begin{tabular}{|C{9mm}|C{9mm}C{8mm}C{9mm}C{9mm}C{10.5mm}|}
\rowcolor[HTML]{89A8F5} 
\hline
Dataset     & Baseline & Bidir.         & Rel.           & +Rem.           & + Rep.       \\ \hline
Web         & 0.514    & 0.548 (3.33\%) & 0.628 (11.36\%) & 0.631 (11.72\%) & \makecell[tc]{0.631 \\ (11.72\%)} \\ \hline
Spr. & 0.308    & 0.308 (0.00\%) & 0.393 (8.49\%)  & 0.429 (12.02\%) & \makecell[tc]{0.429 \\ (12.02\%)} \\ \hline
\end{tabular}}
\caption{On tables without full coverage trans.} \label{tab:bestcov-b}
\end{subtable}
\label{tab:bestcov}
\vspace{-7mm}
\end{table}

\begin{table}[tb]
    \begin{minipage}{.50\linewidth}
\tiny
\centering{\bf
\caption{The number of required \\transformations} \label{tab:reqnum}
\begin{tabular}{|C{9mm}|C{9mm}C{8mm}C{9mm}C{9mm}C{10mm}|}
\rowcolor[HTML]{89A8F5} 
\hline
Dataset     & Baseline & Bidir.         & Rel.           & +Rem.           & + Rep.       \\ \hline
Web         & 431      & 395 (8.35\%)   & 396 (8.12\%)   & 394 (8.58\%)    & \makecell[tc]{394 \\ (8.58\%)}   \\ \hline
Spr. & 970      & 970 (0.00\%)   & 890 (8.25\%)   & 876 (9.69\%)    & \makecell[tc]{876 \\ (9.69\%)}  \\ \hline
\end{tabular}}
    \end{minipage}%
    \hspace{3mm}
    \begin{minipage}{.45\linewidth}
\tiny
\centering{\bf
\caption{The impact of removal and \\repetition concepts}
\label{tab:overall-cov}
\begin{tabular}{|C{9mm}|C{10mm}C{10mm}C{10mm}C{10mm}|}
\hline
\rowcolor[HTML]{89A8F5} 
\cellcolor[HTML]{89A8F5} & \multicolumn{2}{c}{\cellcolor[HTML]{89A8F5}+Rem.} & \multicolumn{2}{c|}{\cellcolor[HTML]{89A8F5}+Rep.} \\ \cline{2-5} 
\rowcolor[HTML]{89A8F5} 
\multirow{-2}{*}{\cellcolor[HTML]{89A8F5}Dataset} & Affected Trans. & Cov. Imp. & Affected Trans. & Cov. Imp. \\ \hline
Web & 20.00\% & 5.90\% & 23.08\% & 6.48\% \\ \hline
Spr. & 8.08\% & 1.56\% & 8.08\% & 1.56\% \\ \hline
\end{tabular}}
    \end{minipage} 
\vspace{-5mm}
\end{table}

\begin{comment}
\begin{table}[t]  
\tiny
\centering{\bf
\caption{The number of required transformations} \label{tab:reqnum}
\begin{tabular}{|C{9mm}|C{9mm}C{8mm}C{9mm}C{9mm}C{10mm}|}
\rowcolor[HTML]{89A8F5} 
\hline
Dataset     & Baseline & Bidir.         & Rel.           & +Rem.           & + Rep.       \\ \hline
Web         & 431      & 395 (8.35\%)   & 396 (8.12\%)   & 394 (8.58\%)    & \makecell[tc]{394 \\ (8.58\%)}   \\ \hline
Spr. & 970      & 970 (0.00\%)   & 890 (8.25\%)   & 876 (9.69\%)    & \makecell[tc]{876 \\ (9.69\%)}  \\ \hline
\end{tabular}}
\end{table}
\end{comment}

\noindent
\textbf{Source-Target Direction Generalization}: The ``bidir.'' column in the aforementioned tables denotes the performance when applying our bidirectional generalization ($\S$~\ref{sec:bidirec-gen}). As demonstrated, our approach outperforms the baseline on the web dataset, which is a relatively more challenging dataset. This shows that the heuristic used in our baseline, which considers columns with lengthier rows as more context-rich, may not accurately label source and target columns. On the other hand, on the spreadsheet benchmark, our performance is on par with the baseline, since in this dataset, the source is always lengthier.%, making the direction detection an easy task for the heuristic approach used in the baseline. 

\noindent
\textbf{Length Generalization}: The utilization of relative indices considerably enhances the performance of our approach compared to the baseline, as observed across all metrics on both datasets. The ``Rel.'' column in Tables~\ref{tab:bestcov} and \ref{tab:reqnum} quantifies this improvement. In each dataset, there are several tables covered with more transformations in the baseline due to the limitations of absolute indices in modeling length variance in the rows. Applying relative indices results in the generation of more general transformations, capable of covering more rows compared to their absolute counterparts. This can lead to up to a 10\% improvement in the coverage of the best transformation and an 8\% reduction in the number of required transformations, particularly on the challenging web dataset.

\noindent
\textbf{Recurrence Generalization}: Columns ``+Rem.'' and ``+Rep.'' denote the performance when augmenting relative-indexed transformations with unit removal and unit removal and repetition, respectively. As shown, there is a slight improvement in both the best transformation coverage and the number of required transformations when employing these generalizations. However, the best transformation coverage and the number of required transformations may not give the full picture when in fact our generalizations considerably enhance the coverage of transformations.
Therefore, instead of relying solely on the best transformation coverage, we measure the improvement for each transformation individually, summarized in Table~\ref{tab:overall-cov}. For this evaluation, we compare the coverage of each transformation when only generalized by relative indices with those that are also augmented by recurrence generalization. Transformations with 100\% coverage and those covering only a single row (i.e., typically literals) are excluded in this evaluation, as they cannot be generalized further. 
As shown, on the web dataset, both unit removal (denoted by ``+Rem.'') and unit removal and repetition (denoted by ``+Rep.'') generalize 20\% or more of the transformations, and the coverage is increased by about 6\%, which underscores the effectiveness and importance of recurrence generalization methods. 

\vspace{-2mm}
\subsection{Generalization to Unseen Data}\label{exp-unseen}%Transforming by Limited Data}
\vspace{-2mm}
The improved coverage on unseen data in our approach is primarily achieved through length and simplicity generalization techniques. 
In our first experiment, we aim to measure the extent to which simplicity generalization can impact the generated transformations on each dataset. Specifically, we conducted the experiment using two tie-breaking strategies when transformations with equal coverage are to be selected. In one strategy, a simple transformation is prioritized over a complex one, and in the other strategy, the selection is made randomly. 
%, as demonstrated in Table~\ref{tab:sim}. 
Our metrics for simplicity in one setting is the total number of units and in another setting is the total number of parameters.
%, respectively denoted by ``Sum unit'' and ``Sum param.'' Table~\ref{tab:sim} shows those quantities as well as the mean number of both units per transformation and parameters per transformation, averaged across all tables of the dataset, respectively denoted by ``Avg unit'' and ``Avg param.'' As shown, 
Simplicity-based tie-breaking results in at least a 5\% reduction in the total number of units and a noticeable 17\% decrease in the parameters needed. Not only do the simpler transformations lead to a faster join on large dynamic environments and provide an easier-to-understand mapping, but, as confirmed by the next experiment, they also yield better coverage on unseen data.

In the next experiment, we randomly divide the rows of each table into two sets. One set represents the available data based on which transformations are built, and the other set represents unseen data that may be added in the future. 
%Assuming that the latter set is gradually incorporated into the table, we construct the transformations on the former set, while the coverage is measured on the entire table. 
We use 10\%, 20\%, and 40\% of the rows as the available data. For tie-breaking among transformations with equal coverage, two strategies are evaluated: simplicity-based and random selection. Table~\ref{tab:limited-random} shows the performance using random selection, and Table~\ref{tab:limited-simple} indicates the same experiment using simplicity-based selection, for both the baseline approach and ours. 

Two important observations can be made:
(1) Regardless of the tie-breaking strategy employed, our approach consistently outperforms the baseline when measuring coverage on unseen data. The performance gap between the baseline and our approach widens when a smaller set of data is available for transformation finding. This implies that, compared to the baseline, our general transformations demonstrate superior performance when generated from a smaller set of available data. 
%As the amount of available data increases, the baseline's performance converges toward our approach in terms of coverage. 
Notably, the gap is more pronounced in the web dataset, reflecting the dataset's inherent complexities.
(2) Employing simplicity-based tie-breaking enhances the coverage of both our approach and the baseline. Nonetheless, our approach benefits more from this generalization, and the improvement is more pronounced in the web dataset, particularly for smaller sample sizes. %For instance, when utilizing 10\% of the available data on the web dataset, the baseline coverage increases by approximately 2\%, whereas our approach demonstrates a 9\% improvement, resulting in a coverage gap of 0.108, which underscores the superior generalization of our approach to unseen data.

\begin{table}[t]
\tiny
\centering
\caption{The impact of generalization on transformation coverage when limited data is available} \label{tab:limited} %Generalization impact on transforming by limited data
\begin{subtable}{.495\textwidth}
\centering{\bf
\begin{tabular}{|C{9mm}|C{10mm}C{10mm}C{18mm}C{8mm}|}
\rowcolor[HTML]{89A8F5} 
\hline
Dataset & Sample & Baseline & General Trans. & Gap \\ \hline
\multirow{3}{*}{Web} & 10\% & 0.528 & 0.590 &0.061\\ 
 & 20\% & 0.663 & 0.704 & 0.041\\ 
 & 40\% & 0.794 & 0.800 & 0.005\\ \hline
\multirow{3}{*}{Spr.} & 10\% & 0.634 & 0.640 & 0.005\\ 
 & 20\% & 0.681 & 0.682 & 0.001\\ 
 & 40\% & 0.793 & 0.801 & 0.008\\ \hline
\end{tabular}}
\caption{Tie-breaking: random} \label{tab:limited-random} %Tie-breaking: random selection
\end{subtable}
\begin{subtable}{.495\textwidth}
\centering{\bf
\begin{tabular}{|C{9mm}|C{10mm}C{10mm}C{18mm}C{8mm}|}
\rowcolor[HTML]{89A8F5} 
\hline
Dataset & Sample & Baseline & General Trans. & Gap \\ \hline
\multirow{3}{*}{Web} & 10\% & 0.537 & 0.645 & 0.108\\
 & 20\% & 0.670 & 0.727 & 0.057\\
 & 40\% & 0.799 & 0.816 & 0.017\\ \hline
\multirow{3}{*}{Spr.} & 10\% & 0.639 & 0.675 & 0.035\\ 
 & 20\% & 0.684 & 0.710 & 0.026\\
 & 40\% & 0.797 & 0.810 & 0.012\\ \hline
\end{tabular}}
\caption{Tie-breaking: simplicity-based} \label{tab:limited-simple} %Tie-breaking: simplicity-based selection
\end{subtable}
\vspace{-7mm}
\end{table}

\vspace{-2mm}
\subsection{End-to-End Join Performance}
\vspace{-2mm}

In the final set of our experiments, we evaluated the performance of our approach on an end-to-end join between two tables with unequal column values. To employ transformations for joinability, we set the repetition degree to one, which ensures a single output for each transformation applied on an input.
%, and adopted the framework proposed in CST, replacing their transformations with our own. 
In addition to the automated similarity-based example generation in CST (referred to as \textit{Automated Example Generation} in the rest of this section), we also experimented with a small set of 5 manually prepared examples for each table, while the remaining rows were kept for testing, denoted as \textit{Manual Example Generation}. Table~\ref{tab:join} summarizes the unequal join performance in terms of Precision, Recall, and F1-Score (denoted with P, R, and F1, respectively) for our approach, CST, and another transformation-based joining approach, Auto-join\footnote{The experiments on Auto-join were conducted using the implementation by the authors of CST~\cite{nobari2022efficiently}.}. Due to the limitations of the sampling framework and subset-based generation in Auto-join, it was only employed when examples were automatically generated and, thus, large enough to be divided into subsets.

As demonstrated, regardless of the sampling method, our approach outperforms all baselines by a noticeable margin in the web dataset, yielding up to an 8\% improvement in F1-Score. The diversity and complexity of the patterns in input rows in this dataset better highlight the ability of our approach to extract more general transformations and, consequently, achieve a better join performance while maintaining the explainability of the mapping space. When examples are automatically generated in the spreadsheet dataset, our performance is on par with the baseline. This is because the dataset is clean, the rows are consistently covered by a few rather easy textual patterns, and the example generation framework returns a sufficient number of correct examples. On the other hand, as shown in Table~\ref{tab:join-samp}, the performance gap exists for both datasets when examples are limited to a few human-generated ones. This gap is expected, considering the better generalization of our approach to unseen data\footnote{Our observations indicate that the best performance-recall trade-off is achieved when the minimum support for transformations is set to 0.05, and all experiments in this section are conducted with that threshold.}.

\begin{table}[tb]
\caption{Join performance of our approach and baselines} \label{tab:join} 
\tiny
    \begin{minipage}{.58\linewidth}    
\centering
\begin{subtable}{\textwidth}
\centering{\bf
\label{tab:join-auto}
\begin{tabular}{|C{3mm}|C{6mm}C{6mm}C{6mm}|C{6mm}C{6mm}C{6mm}|C{6mm}C{6mm}C{6mm}|}
\hline
\rowcolor[HTML]{89A8F5} 
\cellcolor[HTML]{89A8F5} & \multicolumn{3}{c|}{\cellcolor[HTML]{89A8F5}Our Approach} & \multicolumn{3}{c|}{\cellcolor[HTML]{89A8F5}CST} & \multicolumn{3}{c|}{\cellcolor[HTML]{89A8F5}Auto-Join} \\
\rowcolor[HTML]{89A8F5} 
\multirow{-2}{*}{\cellcolor[HTML]{89A8F5}{\rotatebox[origin=c]{90}{}}} & \multicolumn{1}{c}{\cellcolor[HTML]{89A8F5}P} & \multicolumn{1}{c}{\cellcolor[HTML]{89A8F5}R} & \multicolumn{1}{c|}{\cellcolor[HTML]{89A8F5}F1} & \multicolumn{1}{c}{\cellcolor[HTML]{89A8F5}P} & \multicolumn{1}{c}{\cellcolor[HTML]{89A8F5}R} & \multicolumn{1}{c|}{\cellcolor[HTML]{89A8F5}F1} & \multicolumn{1}{c}{\cellcolor[HTML]{89A8F5}P} & \multicolumn{1}{c}{\cellcolor[HTML]{89A8F5}R} & \multicolumn{1}{c|}{\cellcolor[HTML]{89A8F5}F1} \\ \hline
{\rotatebox[origin=c]{90}{Web}} & 0.953 & 0.754 & 0.777 & 0.879 & 0.726 & 0.713 & 0.985 & 0.415 & 0.466 \\ \hline
{\rotatebox[origin=c]{90}{Spr.}} & 0.997 & 0.787 & 0.810 & 0.995 & 0.792 & 0.812 & 0.997 & 0.768 & 0.796 \\ \hline
\end{tabular}}
\caption{Automated example generation} 
\end{subtable}
    \end{minipage} 
    \begin{minipage}{\linewidth}
\begin{subtable}{0.42\textwidth}
\centering{\bf
\begin{tabular}{|C{3mm}|C{6mm}C{6mm}C{6mm}|C{6mm}C{6mm}C{6mm}|}
\hline
\rowcolor[HTML]{89A8F5} 
\cellcolor[HTML]{89A8F5} & \multicolumn{3}{c|}{\cellcolor[HTML]{89A8F5}Our Approach} & \multicolumn{3}{c|}{\cellcolor[HTML]{89A8F5}CST} \\
\rowcolor[HTML]{89A8F5} 
\multirow{-2}{*}{\cellcolor[HTML]{89A8F5}} & \multicolumn{1}{c}{\cellcolor[HTML]{89A8F5}P} & \multicolumn{1}{c}{\cellcolor[HTML]{89A8F5}R} & \multicolumn{1}{c|}{\cellcolor[HTML]{89A8F5}F1} & \multicolumn{1}{c}{\cellcolor[HTML]{89A8F5}P} & \multicolumn{1}{c}{\cellcolor[HTML]{89A8F5}R} & \multicolumn{1}{c|}{\cellcolor[HTML]{89A8F5}F1} \\ \hline
{\rotatebox[origin=c]{90}{Web}} & 0.993 & 0.612 & 0.672 & 0.992 & 0.571 & 0.631  \\ \hline
{\rotatebox[origin=c]{90}{Spr.}} & 0.986 & 0.648 & 0.663 & 0.986 & 0.614 & 0.624 \\ \hline
\end{tabular}}
\caption{Manual example generation} \label{tab:join-samp} %(with 5 examples)
\end{subtable}
    \end{minipage} 
\vspace{-8mm}
\end{table}

\vspace{-2mm}
\section{Conclusion}\label{sec:conclusion} % and Future Works
\vspace{-2mm}
%We have studied the problem of finding general and simple transformations to map joinable columns that are formatted differently. 
%We explored the idea of labelling the source column in each mapping and expanded the domain of transformations by length generalization, removal and repetition concepts. Additionally, we have proposed cluster sampling to address runtime concerns arising from applying generalization concepts. We have conducted both analytical and empirical evaluations of our technique and compared its performance to a state-of-the-art approach. Our evaluations highlight that our method not only produces more efficient transformations but also offers a higher degree of transparency and simplicity compared to competitors.

We have studied the problem of joinablity for non-equi-joinable columns, with a focus on explainability and the generality of transformations. Our exploration has encompassed diverse generalization techniques, such as length invariance, pattern repetition, bi-directionality and simplicity. 
%We have also developed techniques to reduce both the search space and runtime while maintaining the model quality. 
Our evaluation highlights that our method not only produces more effective transformations but also offers a higher degree of simplicity and explainability compared to competitors.
\begin{comment}
Our work can be extended in a few directions:
(1)~Future work may explore different techniques for generating general transformations, other than those discussed in this paper.
(2)~Combining string transformations with the power of large language models while maintaining their explainability is another direction.
(3)~Data integration may be studied under a streaming model where new rows are added to an existing dataset and the previous rows may lose their importance over time.
%(4)~Finally, addressing privacy and security concerns in tabular data integration when combining sensitive data is another direction.
\end{comment}
\vspace{-2mm}
\begin{credits}
\subsubsection{\ackname} This research was supported by the Natural Sciences and Engineering Research Council and the PS752 commemorative scholarship program.
\end{credits}
\vspace{-1mm}

\bibliographystyle{splncs04}
\vspace{-2mm}
\bibliography{main}
\vspace{-2mm}

\end{document}